\documentclass[aps,superscriptaddress,groupedaddress]{revtex4}
\usepackage{graphicx}
\usepackage{dcolumn}
\usepackage{bm}
\usepackage{amssymb}
\usepackage{amsfonts}
\usepackage{amsmath}

\begin{document}

\title[Sequential generation...]{Sequential generation of Polynomial
Invariants and N-body non-local correlations }
\author{S. Shelly Sharma}
\email{shelly@uel.br}
\affiliation{Departamento de F\'{\i}sica, Universidade Estadual de Londrina, Londrina
86051-990, PR Brazil }
\author{N. K. Sharma}
\email{nsharma@uel.br}
\affiliation{Departamento de Matematica, Universidade Estadual de Londrina, Londrina
86051-990, PR Brazil }

\begin{abstract}
We report an inductive process that allows for sequential construction of
local unitary invariant polynomials of state coefficients for multipartite
quantum states. The starting point can be a physically meaningful invariant
of a smaller part of the system. The process is applied to construct a chain
of invariants that quantify non-local $N-$way correlations in an $N-$qubit
pure state. It also yields the invariants to quantify the sum of $N-$way and 
$\left( N-1\right) $-way correlations. Analytic expressions for four-way and
three-way correlation quantifiers for four qubit states, as well as,
five-way and four-way correlation quantifiers for five qubit pure states are
given.
\end{abstract}

\maketitle

\section{Introduction}

Quantum entanglement, regarded as one of the most prominent features of
quantum mechanics, is a nonlocal property of a quantum state determined by
quantum correlations present in the system. Entanglement is a necessary
ingredient of any quantum computation and a physical resource for quantum
cryptography and quantum communication. The negativity \cite{zycz98,vida02}
of partial transpose \cite{pere96} of the state operator was introduced to
detect and measure entanglement of bipartite states. Multipartite
entanglement that comes into play in quantum systems with more than two
subsystems, is a resource for multiuser quantum information tasks. Since the
mathematical structure of multipartite states is much more complex than that
of bipartite states, the classification of multipartite entanglement is a
far more challenging task.

It is natural to expect that the intrinsic non-local correlations in a
multipartite system are a function of non-local correlations present in
constituent subsystems. Therefore, polynomial functions of state
coefficients that remain invariant under local unitary operations are used
to quantify non-local correlations present in the state. Local unitary
invariance implies that the value of invariant is independent of the choice
of local basis for the subsystem. Local unitary invariant polynomial
functions of state coefficients are known to discriminate between and
quantify distinct entanglement types as in the case of three-qubit states 
\cite{sudb01}. Relations between polynomial invariants and entanglement have
been investigated in \cite{lind98,cart99,rain00,gras98}. In this article, we
report a general inductive process which allows to sequentially generate
polynomial unitary invariants for a multipartite quantum state from known
subsystem invariants. The process is described in the context of qubit
systems but can be generalized to multilevel systems with $d>2$. To examine
degree $k$ functions of state coefficients of an $N-$qubit state, we work in
the state space of $k$ copies of $N-$qubit state. If \ a degree $k$ function
of state coefficients is an $\left( N-1\right) $ qubit invariant then a
unitary transformation on $N^{th}$ qubit results in a binary form.
Invariants of the binary form are easily obtained by using standard methods
of classical invariant theory. The physical meaning of set of polynomial
invariants generated through the process can be inferred from the subsystem
invariant which is selected as the starting point. Main result of the
article is that an $N$-way tangle constructed from degree $2k$ invariant is
a function of degree $k$ invariants that quantify ($N-1$)-way correlations
in an $N$-qubit state. A weighted sum of squares of ($N-1$)-way tangles
determines the total amount of $N$-way and ($N-1$)-way quantum correlations
present in the pure state. An $N-$way tangle is a computable measure of
genuine $N-$way entanglement of an $N-$qubit pure state.

Luque and Thibon \cite{luqu03, luqu06} have used classical invariant theory
to obtain unitary polynomial invariants for four qubit systems. The
invariants got geometrical meaning in the work of Levay \cite{leva05,leva06}%
. Other interesting efforts to calculate the polynomial invariants of four
and more qubits include refs. \cite%
{wong01,leif04,oste06,luqu07,zoko09,dli09,shar101,shar102,vieh11,xli12,
shar12,elts12,xli13}. More recently, algebraic geometry and invariant theory
have been used to investigate the entanglement classes of four qubit states 
\cite{holw14}, and relationship between invariants for qubits and invariants
for spinors has been discussed in ref. \cite{leva15}. Usual approach is to
look for all the polynomial invariants for a given number of qubits. As the
number of local unitary invariants grows exponentially with the number of
constituent subsystems, it becomes a formidable task to identify physically
meaningful invariants out of a maze of invariants. Using our process, one
can focus on construction of invariants that detect and quantify a specific
property of a multipartite system and the constituent subsystems in a
quantum state. In particular, the polynomial invariants that have physical
interpretation in terms of entanglement are of great interest as these may
be used to quantify the entanglement resource content of a pure state. It is
known that local unitaries cannot link a state on which a given polynomial
invariant is non-zero to another state on which the value of same invariant
is null. Selective sequential generation of polynomial invariants, proposed
in this article, makes it possible to group together states with a specific
type of entanglement.

To illustrate the effectiveness of our inductive process we describe the
sequence of steps to construct polynomial invariants that detect and
quantify $N-$way non-local correlations. Our starting point is a degree two
polynomial invariant, which is the negative eigenvalue of partially
transposed two qubit pure state. Global negativity \cite{zycz98,vida02} or
equivalently concurrence \cite{hill97} of a two qubit pure state in known to
quantify two-way correlations. Three tangle \cite{coff00} of a three qubit
pure state, is an entanglement monotone that quantifies three-way
correlations. Four -way correlations in a four qubit pure state are
quantified by four-tangle \cite{shar13}, a function of degree eight
polynomial invariant. Three tangle and four tangle are moduli of specific
local unitary invariants that are polynomial functions of state
coefficients. The process allows us to relate global negativity, three
tangle, and four-tangle to five tangle and higher degree tangles. Exact
mathematical relation between $N-$way polynomial invariants and $\left(
N-1\right) -$way invariants of a quantum state is found.

In section II, we discuss the structure of $\left( N-1\right) $ qubit
invariants of an $N$ qubit pure state. An index raising operator that
relates the elements in the set of $\left( N-1\right) $ qubit invariants is,
also, defined. Step by step construction of three-qubit invariants of four
qubit state, by successive applications of index raising operator, is given
in Appendix A. The formalism to construct $N-$qubit invariants in terms of $%
\left( N-1\right) $ qubit invariants is given in section III. Appendix B is
a note on transvection process to obtain invariants of a binary form. The
principal construction tools that is local unitary (LU) invariance,
selective partial transposition \cite{shar07,shar08} and notion of
negativity fonts, used in our earlier works \cite{shar101,shar102}, are also
needed for sequential construction of polynomial invariants that quantify $N$%
-way correlations. For the sake of completeness, we define the determinants
of negativity fonts in section IV. In section V, we generate polynomial
invariants with negativity of partial transpose of a two qubit pure state as
the starting point. To illustrate the process, an invariant that can
quantify genuine five-way entanglement of five qubit state is written down
and five-tangle defined. Although some five qubit invariants have been
reported in ref. \cite{luqu06}, an analytical form in terms of state
coefficients for a degree 16 invariant that detects genuine five-way
correlations is being reported for the first time. A summary of results is
in section VI.

\section{Structure of $\left( N-1\right) $ qubit invariants of an $N$ qubit
state}

A quantum system of $N$ isolated qubits in a pure state is described by a
wave-function%
\begin{equation}
\left\vert \Psi ^{A_{1}A_{2}...A_{N}}\right\rangle
=\sum_{i_{1}i_{2}...i_{N}}a_{i_{1}i_{2}...i_{N}}\left\vert
i_{1}i_{2}...i_{N}\right\rangle ,  \label{nqubitstate}
\end{equation}%
where $\{\left\vert e_{\mu }\right\rangle =\left\vert
i_{1}i_{2}...i_{N}\right\rangle :\mu =1,...,d\}$ is the set of basis vectors
spanning a Hilbert space $\mathcal{H}=\left( 
\mathbb{C}
^{2}\right) ^{\otimes N}$ of dimension $d=2^{N}$, and $A_{q}$ is the
location of qubit number $q$. The coefficients $a_{i_{1}i_{2}...i_{N}}$ are
complex numbers. The local basis states of a single qubit are labelled by $%
i_{q}=0$ and $1,$ where $q=1,2,...,N$. How does a degree $k$ function of
state coefficients transform under a local unitary? The state space of $k$
copies of the state, $\mathcal{H}^{\otimes k},$ is spanned by vectors in the
set $B^{k}=\left\{ \left\vert e_{\mu _{1}}\right\rangle \otimes ...\otimes
\left\vert e_{\mu _{k}}\right\rangle :\mu _{i}=1,d\right\} $. We may write $%
B^{k}=\cup _{m=0}^{k}B_{q}^{k-m}$, where the subset $B_{q}^{k-m}$ is defined
as $B_{q}^{k-m}$=$\left\{ \left\vert e_{\mu _{1}}\right\rangle \otimes
...\otimes \left\vert e_{\mu _{k}}\right\rangle :0_{q}^{\otimes k-m}\otimes
1_{q}^{\otimes m},\mu _{i}=1,d\right\} $, with $m=0,...,k$. Subset $%
B_{q}^{k-m}$ groups together those vectors in $\mathcal{H}^{\otimes k}$ in
which $\left\vert i_{q}=0\right\rangle $ appears $k-m$ times and $\left\vert
i_{q}=1\right\rangle $ appears $m$ times. The number of elements in subset $%
B_{q}^{k-m}$ is given by the binomial coefficient defined as $\binom{k}{m}=%
\frac{k!}{m!\left( k-m\right) !}$. In the expansion of $\left\vert \Psi
^{A_{1}A_{2}...A_{N}}\right\rangle ^{\otimes k}$ in the basis $B^{k}$,
coefficients of elements that belong to subset $B_{q}^{k-m}$ transform under
a unitary on qubit $q$ as the coefficients of $\left\vert 0^{\otimes
k-m}\otimes 1^{\otimes m}\right\rangle $.

We now examine how the coefficients in product of $k$ copies of the state
transform under a unitary transformation on $N^{th}$ qubit. To simplify the
notation, we represent a state coefficient in $\nu ^{th}$ copy of $%
\left\vert \Psi ^{A_{1}A_{2}...A_{N}}\right\rangle $ by $a_{\left(
t_{N-1},\nu \right) \left( i_{N,\nu }\right) }$ where $t_{N-1}\equiv
i_{1}i_{2}...i_{N-1}$. Consider a degree $k$ monomial of state coefficients 
\begin{equation*}
f_{k,0}=\prod\limits_{\nu =1}^{k}a_{\left( t_{N-1},\nu \right) \left(
0,v\right) }.
\end{equation*}%
If the state transforms under a local unitary, $U^{A_{N}}=\frac{1}{\sqrt{%
1+\left\vert x\right\vert ^{2}}}\left[ 
\begin{array}{cc}
1 & -x^{\ast } \\ 
x & 1%
\end{array}%
\right] ,$as%
\begin{eqnarray*}
\left\vert \Psi ^{A_{1}A_{2}...A_{N}}\right\rangle &=&\frac{1}{\sqrt{%
1+\left\vert x\right\vert ^{2}}}\sum_{t_{N-1}}\left( a_{t_{N-1}0}-x^{\ast
}a_{t_{N-1}1}\right) \left\vert t_{N-1}0\right\rangle \\
&&+\frac{1}{\sqrt{1+\left\vert x\right\vert ^{2}}}\sum_{t_{N-1}}\left(
a_{t_{N-1}1}+xa_{t_{N-1}0}\right) \left\vert t_{N-1}1\right\rangle ,
\end{eqnarray*}%
then $f_{k,0}$ transforms as coefficient of a vector in subset $B_{N}^{k}$
that is the transformed function is%
\begin{eqnarray}
f_{k,0}^{\prime } &=&\frac{1}{\left( 1+\left\vert x\right\vert ^{2}\right)
^{k/2}}\prod\limits_{\nu =1}^{k}\left( a_{\left( t_{N-1},\nu \right) \left(
0,\nu \right) }-x^{\ast }a_{\left( t_{N-1},\nu \right) \left( 1,\nu \right)
}\right) \text{.}  \notag \\
&=&\frac{1}{\left( 1+\left\vert x\right\vert ^{2}\right) ^{k/2}}%
\sum\limits_{m=0}^{k}\left( -x^{\ast }\right) ^{m}\binom{k}{m}f_{k-m,m}.
\label{nthqubit}
\end{eqnarray}%
The functions $f_{k-m,m}$ ($m=0$ to $k$) can be obtained from $f_{k,0}$ by
the action of an index raising operator $T_{A_{N}}^{+}$ defined as%
\begin{equation}
T_{A_{N}}^{+}a_{t_{N-1}0}=a_{t_{N-1}1},\quad T_{A_{N}}^{+}a_{t_{N-1}1}=0.
\label{rone}
\end{equation}%
On a product of state coefficients the operator acts as follows: 
\begin{equation}
T_{A_{N}}^{+}\left( a_{t_{N-1}i_{N}}a_{t_{N-1}^{\prime }i_{N}^{\prime
}}\right) =\left( T_{A_{N}}^{+}a_{t_{N-1}i_{N}}\right) a_{t_{N-1}^{\prime
}i_{N}^{\prime }}+a_{t_{N-1}i_{N}}\left( T_{A_{N}}^{+}a_{t_{N-1}^{\prime
}i_{N}^{\prime }}\right) .  \label{rtwo}
\end{equation}%
With index raising operator defined as above, it is easily verified that%
\begin{equation*}
f_{k-m,m}=\frac{\left( k-m\right) !}{k!}\left( T_{A_{N}}^{+}\right)
^{m}f_{k,0}.
\end{equation*}%
The function $f_{k-m,m}$ is a function of coefficients of elements in $%
B_{N}^{k-m}$, as such transforms as coefficient of $\left\vert
0_{N}^{\otimes k-m}\otimes 1_{N}^{\otimes m}\right\rangle $.

Next, let $I_{N-1,k}$ be a homogeneous degree $k$ polynomial invariant of $%
\left( N-1\right) $ ($N>2$) qubit system. Corresponding $\left( N-1\right) $
qubit invariant of $N$ qubit state\ that transforms as coefficient of $%
\left\vert 0^{\otimes k}\right\rangle $ ($\left\vert 1^{\otimes
k}\right\rangle $)\ is denoted by $\left( I_{N-1}\right) _{A_{N}}^{k,0}$ ($%
\left( I_{N-1}\right) _{A_{N}}^{0.k}$) and is readily written down by
substituting the state coefficients $a_{t_{N-1}}$ by $a_{t_{N-1}0}$ ($%
a_{t_{N-1}1}$). Superscript $\left( k-m,m\right) $ means that every term in
the expansion of $\left( I_{N-1}\right) _{A_{N}}^{k-m,m}$ is a product of $k$
state coefficients such that exactly $\left( k-m\right) $ coefficients have $%
i_{N}=0$, while $m$ state coefficients have $i_{N}=1$. The $\left(
N-1\right) $ qubit invariant $\left( I_{N-1}\right) _{A_{N}}^{k-m,m},$ is a
function of coefficients of elements in $B_{N}^{k-m}$. By linearity, $\left(
I_{N-1}\right) _{A_{N}}^{k-m,m}$ transforms under a unitary on N$^{th}$
qubit, in a manner analogous to that of functions $f_{k-m,m}$. In analogy
with Eq. (\ref{nthqubit}), transformation equation for $\left(
I_{N-1}\right) _{A_{N}}^{k,0}$\ under a local unitary $U^{A_{N}}$ on qubit $%
A_{N}$, reads as%
\begin{equation}
\left( I_{N-1}^{\prime }\right) _{A_{N}}^{k,0}=\frac{1}{\left( 1+\left\vert
x\right\vert ^{2}\right) ^{k/2}}\sum\limits_{m=0}^{k}\left( -x^{\ast
}\right) ^{m}\binom{k}{m}\left( I_{N-1}\right) _{A_{N}}^{k-m,m}.
\label{kmin}
\end{equation}%
The invariants $\left( I_{N-1}\right) _{A_{N}}^{k-m,m}$ for $1<m<\left(
k-1\right) $ are additional ($N-1$) qubit invariants of $\left\vert \Psi
_{N}\right\rangle $ with no counterparts in $N-1$ qubit state. These can be
constructed from $\left( I_{N-1}\right) _{A_{N}}^{k,0}$ through the action
of raising operator $T_{A_{N}}^{+}$ on state coefficients that appear in the
analytical form of $\left( I_{N-1}\right) _{A_{N}}^{k,0}$. Using the
definition of $T_{A_{N}}^{+}$ (Eqs. (\ref{rone} and \ref{rtwo})) one can
verify that 
\begin{equation}
\left( T_{A_{N}}^{+}\right) ^{m}\left( I_{N-1}\right) _{A_{N}}^{k,0}=\frac{k!%
}{\left( k-m\right) !}\left( I_{N-1}\right) _{A_{N}}^{k-m,m},  \label{tp_inv}
\end{equation}%
where the structure of $I_{N-1,k}$ determines the analytical expression of $%
\left( I_{N-1}\right) _{A_{N}}^{k-m,m}$ in terms of state coefficients. If V$%
^{k+1}$ is the vector space of $N-1$ qubit invariants of degree $k$ then
elements of the set of invariants $\left\{ \left( I_{N-1}\right)
_{A_{N}}^{k-m,m}:0\leq m\leq k\right\} $ form a coordinate ring. An element $%
\left( I_{N-1}\right) _{A_{N}}^{k-m,m}$ of the ring transforms, under a
local unitary on $N^{th}$ qubit, as the coefficient of $\left\vert
0^{\otimes k-m}1^{\otimes m}\right\rangle $. As such a local unitary on $%
N^{th}$ qubit may be used to transform a given $N$ qubit state to a form
such that one of the elements in the set $\left\{ \left( I_{N-1}^{\prime
}\right) _{A_{N}}^{k-m,m}:0\leq m\leq k\right\} $ is zero. For example, the
unitary that makes $\left( I_{N-1}^{\prime }\right) _{A_{N}}^{k,0}=0$, is
determined by the condition 
\begin{equation}
P\left( x^{\ast }\right) =\sum\limits_{m=0}^{k}\binom{k}{m}\left( -x^{\ast
}\right) ^{m}\left( I_{N-1}\right) _{A_{N}}^{k-m,m}=0,  \label{kminp}
\end{equation}%
which is a degree $k$ polynomial in variable $x^{\ast }$ with $N-1$ qubit
invariants as coefficients. Invariants of the polynomial of Eq. (\ref{kminp}%
) are the $N$ qubit invariants we are looking for.

In ref. \cite{gaur13}, homogenous polynomials of a fixed degree $k$, are
seen as vectors in a Hilbert space consisting of $k-$copies of the original
Hilbert space and a technique to calculate all SL-invariant polynomials of
degree $k$ in any number of qudits is given. In our construction, on the
other hand, the structure of an invariant polynomial of degree $2k$ for $N$
qubits is determined by the structure of degree $k$ invariant polynomials of 
$\left( N-1\right) $ qubits of the same state. Therefore the physical
meaning of invariants of Eq. (\ref{kminp}) depends on the property of the $%
N- $qubit state that the underlying $\left( N-1\right) $ qubit invariants
represent. For instance, if finite $I_{N-1,k}$ represents genuine $\left(
N-1\right) $ qubit entanglement then the entanglement of a selected set of $%
\left( N-1\right) $ qubits in N-qubit state is bound to be a function of
elements in the set of invariants $\left\{ \left( I_{N-1}\right)
_{A_{N}}^{k-m,m},m=0,k\right\} $.

\section{Polynomial Unitary invariants of N qubits and Binary form}

Comparing polynomial $P\left( x^{\ast }\right) $ of Eq. (\ref{kminp}) with
binary form%
\begin{equation}
f\left( x_{1},y_{1}\right) =\sum\limits_{\mu =0}^{k}\binom{k}{\mu }\left(
x_{1}\right) ^{\mu }\left( y_{1}\right) ^{k-\mu }\left( I_{N-1}\right)
_{A_{N}}^{k-\mu ,\mu },
\end{equation}%
we notice that $P\left( x^{\ast }\right) =f\left( -x^{\ast },1\right) $.
Invariants of the binary form are the same as $N-$ qubit invariants of the
polynomial. As explained in Appendix (B), the standard method of
transvection may be used to obtain all covariants and invariants of the
binary form, which are also covariants and invariants of the corresponding
polynomial. The number of polynomial invariants of a binary form depends on
the order of the form. Since in Eq. (\ref{kmin}) coefficients $\left(
I_{N-1}\right) _{A_{N}}^{k-\mu ,\mu }$ are degree $k$ functions of state
coefficients, the degree of $N-$qubit invariants obtained from Eq. (\ref%
{kmin}) depends on $k$ as well as order of the form. To illustrate the power
of the process, we construct in section V the sequence of invariants based
on negativity of partial transpose of a two qubit pure state. Global
negativity \cite{zycz98,vida02} of a bipartite state is a well known
entanglement monotone and in the case of two qubits quantifies two-way
non-local correlations. We are mainly interested in polynomial invariants
relevant to quantifying $N-$way and $N-1$-way correlations in an $N-$qubit
state. In this context, we focus on two specific invariants of Eq. (\ref%
{kmin}), which are shown in Appendix (B) to have the form%
\begin{equation}
I_{N,2k}=\frac{1}{2}\sum\limits_{m=0}^{k}\left( -1\right) ^{m}\binom{k}{m}%
\times \left( I_{N-1}\right) _{A_{N}}^{k-m,m}\left( I_{N-1}\right)
_{A_{N}}^{m,k-m}  \label{inv1}
\end{equation}%
and%
\begin{equation}
\left( \mathcal{N}_{N,2k}\right) ^{A_{1}A_{2}...A_{N-1}\left( A_{N}\right)
}=\sum\limits_{m=0}^{k}\binom{k}{m}\left\vert \left( I_{N-1}\right)
_{A_{N}}^{k-m,m}\right\vert ^{2}.  \label{inv2}
\end{equation}%
Since $N-1$ qubits including qubit $A_{1}$ may be selected in $N-1$ distinct
ways, and the value of invariant of Eq. (\ref{inv2}) depends on this choice.
We define $\left( \mathcal{N}_{N,2k}\right)
^{A_{1}A_{2}...A_{q-1}A_{q+1}...A_{N}\left( A_{q}\right) }$ to be the
invariant of second type when $N-1$ qubits including A$_{1}$ do not include
qubit $A_{q},$ ($2\leq q\leq N)$ and construct an overall invariant, $%
\mathcal{N}_{N,2k}^{A_{1}}$, which is a normalized sum of invariants of the
second type corresponding to each selection and reads as%
\begin{eqnarray}
\mathcal{N}_{N,2k}^{A_{1}} &=&C_{N}\left( \mathcal{N}_{N,2k}\right)
^{A_{1}A_{2}...A_{N-1}\left( A_{N}\right) }  \notag \\
&&+C_{N}\sum_{q=2}^{N-1}\left( \mathcal{N}_{N,2k}\right)
^{A_{1}A_{2}...A_{q-1}A_{q+1}...A_{N}\left( A_{q}\right) }.  \label{inv3}
\end{eqnarray}%
Here $C_{N}$ is the normalization constant chosen so that $0\leq \mathcal{N}%
_{N,2k}^{A_{1}}\leq 1.$

The invariant $I_{N,2k}$ is a combination of terms which are products of $%
N-1 $ qubit invariants, therefore, if $I_{N-1,k}$ is an invariant that
detects genuine $\left( N-1\right) -$way correlations then $I_{N,2k}$
detects genuine $N-$way correlations. It is obviously zero on all $N$ qubit
states, with no $N-$way or $\left( N-1\right) $ way correlations. We can
verify that the invariant $I_{N,2k}$ is zero on any pure state which is a
product state of an ($N-1$) qubit state with nonzero $N-1$ qubit
correlations and a single qubit state. For example, on a state for which $%
N^{th}$ qubit is separable that is%
\begin{equation}
\left\vert \Phi _{1-s}^{A_{1}A_{2}...A_{N}}\right\rangle =\left(
\sum_{i_{1}i_{2}...i_{N-1}}a_{i_{1}i_{2}...i_{N-1}}\left\vert
i_{1}i_{2}...i_{N-1}\right\rangle \right) \left( \alpha _{0}\left\vert
0\right\rangle +\alpha _{1}\left\vert 1\right\rangle \right) ,  \label{sep}
\end{equation}%
the product%
\begin{equation}
\left( I_{N-1,k}\right) _{A_{N}}^{k-m,m}\left( I_{N-1,k}\right)
_{A_{N}}^{m,k-m}=\left( I_{N-1,k}\right) ^{2}\left( \alpha _{0}\right)
^{k}\left( \alpha _{1}\right) ^{k}\text{.}
\end{equation}%
By using the properties of Binomial coefficients we obtain%
\begin{equation}
I_{N,2k}\left( \left\vert \Phi _{1-s}^{A_{1}A_{2}...A_{N}}\right\rangle
\right) =\frac{1}{2}\left( I_{N-1,k}\right) ^{2}\left( \alpha _{0}\right)
^{k}\left( \alpha _{1}\right) ^{k}\sum\limits_{m=0}^{k}\left( -1\right) ^{m}%
\binom{k}{m}=0.
\end{equation}%
It is easily verified that the value of invariant $I_{N,2k}$ does not depend
on the choice of $N-1$ qubits, that the invariant $I_{N-1,k}$ refers to.

The local unitary group G on $N-$qubit states is the group of 1-qubit
operations. G can be taken to be G $=$ U(1) $\times $ SU(2)$^{N}$ in the
case of pure states. Explicitly, for pure states, an element $g=(e^{i\theta
},g_{1},...,g_{N})\in $ G acts on $\left\vert \Psi \right\rangle $ by 
\begin{equation*}
g\left\vert \Psi \right\rangle =e^{i\theta }\left( g_{1}\otimes ...\otimes
g_{N}\right) \left\vert \Psi \right\rangle .
\end{equation*}%
By construction $I_{N,2k}$ of Eq. (\ref{inv1}) is invariant with respect to
SU(2)$^{N}$, but not with respect to U(1). However $\left\vert
I_{N,2k}\right\vert $ and $\mathcal{N}_{N,2k}^{A_{1}}$ are invariant with
respect to U(1) $\times $ SU(2)$^{N}$.

As shown in section V, with negativity of partially transposed state
operator of a two qubit pure state as a starting point, the process outlined
above yields polynomial invariants to detect $3-$way, $4-$way, ..., $N-$way
non-local correlations. In order to use Eqs. ((\ref{kmin}),(\ref{inv1}), and
(\ref{inv2})) for a given $N-$qubit state, we need the complete set of $%
(N-1) $ qubit invariants $\left( I_{N-1}\right) _{A_{N}}^{k-m,m}$ $\left(
0\leq m\leq k\right) $. Section IV provides the elements to construct $(N-1)$
qubit invariants in terms of determinants of negativity fonts, which are
single qubit invariant functions of state coefficients.

\section{Determinant of a $K-$way Negativity Font}

Construction of \ a sequence of polynomial invariants requires obtaining the
set of ($N-1$) qubit invariants for a given $N$-qubit system. To construct
invariants based on negativity of partial transpose of a two qubit pure
state it is useful to express the invariants involved at different stages of
the process in terms of determinants of two by two matrices of state
coefficients. Each matrix, referred to as a negativity font, determines the
negative eigenvalue due to partial transposition with respect to a focus
qubit in a specific four by four subspace of Hilbert space of the system.
Determinant of a given negativity font detects the potential entanglement
present in a specific four by four subspace of Hilbert space $\mathcal{H}$.
In this sense, negativity fonts can be seen as the elementary units of
entanglement in a quantum superposition. By construction, the determinant of
the matrix is a degree two function which is invariant with respect to a
unitary transformation on the focus qubit. This section contains a formal
definition of determinant of a negativity font, followed by its
transformation under the action of a raising operator defined in Eq. (\ref%
{rone}).

Consider the state operator of $N$ qubit pure state (Eq. (\ref{nqubitstate}%
)), $\rho =\left\vert \Psi ^{A_{1}A_{2}...A_{N}}\right\rangle \left\langle
\Psi ^{A_{1}A_{2}...A_{N}}\right\vert $. Let $\rho ^{T_{A_{1}}}$ represent
the partially transposed state operator obtained from $\rho $ by
transposition with respect to focus qubit $A_{1}.$ Using the symbol $\oplus $
to represent addition modulo $2$, a negativity font that belongs to $\rho
^{T_{A_{1}}}$ is given by a two by two matrix%
\begin{equation}
\nu _{S_{2,T}}^{i_{1}i_{2}...i_{N}}=\left[ 
\begin{array}{cc}
a_{i_{1}i_{2}...i_{N}} & a_{j_{1}\oplus 1,j_{2}...j_{N}} \\ 
a_{i_{1}\oplus 1,i_{2}...i_{N}} & a_{j_{1}j_{2}...j_{N}}%
\end{array}%
\right] .  \label{nfont}
\end{equation}%
We next explain the meaning of subscript $S_{2,T}$ in $\nu
_{S_{2,T}}^{i_{1}i_{2}...i_{N}}$. The set of $N$ qubits with their locations
and local basis indices given by, $T=\left\{ \left( A_{1}\right)
_{i_{1}},\left( A_{2}\right) _{i_{2}}...\left( A_{N}\right) _{i_{N}}\right\}
,$ is split up into two subsets. Subset $S_{1,T}$ contains $K$ qubits with $%
i_{m}\neq j_{m}$ in Eq. (\ref{nfont})), and $S_{2,T}$ contains the remaining 
$N-K$ qubits with $i_{m}=j_{m}$ . More specifically, $\nu
_{S_{2,T}}^{i_{1}i_{2}...i_{N}}$ is referred to as a $K-$way negativity font 
\cite{shar13} , where $K=\sum\limits_{m=1}^{N}\left( 1-\delta
_{i_{m}j_{m}}\right) $ ($\delta _{i_{m}j_{m}}=0$ for $i_{m}\neq j_{m}$ and $%
\delta _{i_{m}j_{m}}=1$ for $i_{m}=j_{m}$). The determinant of a negativity
font is defined as $D_{S_{2,T}}^{s_{1,T}}=\det \left( \nu
_{S_{2,T}}^{i_{1}i_{2}...i_{N}}\right) $, where $s_{1,T}$ represents the
sequence of local basis indices for qubits in set $S_{1,T}$. We notice that
any $D_{S_{2,T}}^{s_{1,T}}$ is invariant with respect to a unitary on qubit $%
A_{1}$ while the determinant given by $D_{S_{2,T}}^{00}$ is a two qubit
invariant with respect to unitaries on qubits that appear in the
superscript. If a local unitary is applied to qubit $\left( A_{p}\right)
_{i_{p}}$ ($p\neq 1$) in set $S_{2,T}$, then the determinant $%
D_{S_{2,T}}^{s_{1,T}}$ transforms as the coefficient of $\left\vert \left(
i_{p}\right) ^{\otimes 2}\right\rangle $. However, if qubit $\left(
A_{p}\right) _{i_{p}}$ ($p\neq 1$) belongs in set $S_{1,T}$ then $%
D_{S_{2,T}}^{s_{1,T}}$ transforms as coefficient of $\left\vert 0\otimes
1\right\rangle $. Negativity fonts and their determinants that quantify the
negativity of $\rho ^{T_{A_{p}}}$ for $p\neq 1$ are defined in an analogous
fashion.

Determinants of $K-$way negativity fonts are linked to $K-$way partially
transposed matrices obtained from state operator through selective partial
transposition \cite{shar07,shar08}. In refs. \cite{shar101,shar102}, the
concept of negativity fonts was used to construct invariants that quantify
three and four-way correlations for three and four qubit pure states,
respectively. In a multipartite state having $N$ subsystems, entanglement of
a given subsystem to its complement, may arise due to $2-$way, $3-$way, ...,$%
N-$way correlations. Since the most general multiqubit state may have $N-1$
correlation types, it can have as many types of negativity fonts. Local
unitary transformations on qubits result in a state with a new set of
negativity fonts. Unitary invariant polynomial functions of state
coefficients that are finite on states having a specific entanglement type,
are functions of determinants of negativity fonts that remain invariant
under local unitary transformations.

\subsection{Action of a raising operator on determinant of Negativity Font}

How does the index raising operator relate the determinants of $N$ and $%
\left( N-1\right) $ qubit negativity fonts? Let $D_{S_{2,T}}^{s_{1,T}}=\det
\left( \nu _{S_{2,T}}^{i_{1}i_{2}...i_{N-1}}\right) $ be the determinant of
an arbitrary negativity font in an $\left( N-1\right) $ qubit state, for
partial transposition with respect to qubit $A_{1}$. Two $N$ qubit
determinants obtained by adding N$^{th}$ qubit that is subscript $\left(
A_{N}\right) _{i_{N}}$ to set $S_{2,T}$ are%
\begin{equation}
D_{\left( S_{2,T}\right) \left( A_{N}\right) _{i_{N}}}^{s_{1,T}}=\det \left(
\nu _{\left( S_{2,T}\right) \left( A_{N}\right)
_{i_{N}}}^{i_{1}i_{2}...i_{N}}\right) ;\qquad i_{N}=0,1.
\end{equation}%
Determinants $D_{\left( S_{2,T}\right) \left( A_{N}\right)
_{i_{N}}}^{s_{1,T}}$ transform under local unitaries as coefficient of $%
\left\vert i_{N}^{\otimes 2}\right\rangle $. Two distinct determinants
obtained by adding $i_{N}$ to set $s_{1,T}$ transform as coefficients of $%
\left\vert 0\otimes 1\right\rangle $. The combination 
\begin{equation}
\frac{1}{2}\left( D_{S_{2,T}}^{\left( s_{1,T}\right) 0}-D_{S_{2,T}}^{\left(
s_{1,T}\right) 1}\right) =\frac{1}{2}\left( \det \left( \nu
_{S_{2,T}}^{i_{1}i_{2}...i_{N}=0}\right) -\det \left( \nu
_{S_{2,T}}^{i_{1}i_{2}...i_{N}=1}\right) \right)
\end{equation}%
is invariant with respect to local unitaries on qubits $A_{1}$ and $A_{N}$,
whereas the sum $\frac{1}{2}\left( D_{S_{2,T}}^{\left( s_{1,T}\right)
0}+D_{S_{2,T}}^{\left( s_{1,T}\right) 1}\right) $ is not so. It is easily
verified that%
\begin{equation}
T_{A_{N}}^{+}D_{\left( S_{2,T}\right) \left( A_{N}\right)
_{0}}^{s_{1,T}}=\left( D_{S_{2,T}}^{\left( s_{1,T}\right)
0}+D_{S_{2,T}}^{\left( s_{1,T}\right) 1}\right) ,  \label{tp_det_sub}
\end{equation}%
\begin{equation}
T_{A_{N}}^{+}\left( D_{S_{2,T}}^{\left( s_{1,T}\right)
0}+D_{S_{2,T}}^{\left( s_{1,T}\right) 1}\right) =2D_{\left( S_{2,T}\right)
\left( A_{N}\right) _{1}}^{s_{1,T}},  \label{tp_det_sup}
\end{equation}%
\begin{equation}
T_{A_{N}}^{+}D_{\left( S_{2,T}\right) \left( A_{N}\right) _{1}}^{s_{1,T}}=0.
\end{equation}%
On a product of determinants of negativity fonts, $D_{1}D_{2}$, we have 
\begin{equation}
T\left( D_{1}D_{2}\right) =\left( TD_{1}\right) D_{2}+D_{1}\left(
TD_{2}\right) .  \label{tp_det_prod}
\end{equation}

\section{Polynomial invariants of an N qubit state based on negativity of a
two qubit pure state}

In this section, the process outlined in sections II and III is applied to
generate the sequence of polynomial invariants from which one can construct
entanglement monotones to quantify three-way, four-way and five-way
correlations in three, four and five qubit states. A two qubit pure state
shared by Alice (qubit $A_{1}$) and Bob (qubit $A_{2}$),%
\begin{equation}
\left\vert \Psi ^{A_{1}A_{2}}\right\rangle
=\sum_{i_{1}i_{2}}a_{i_{1}i_{2}}\left\vert i_{1}i_{2}\right\rangle
,i_{m}=0,1,
\end{equation}%
has a single two-way negativity font with determinant $D^{00}=\left(
a_{00}a_{11}-a_{10}a_{01}\right) $. It is a degree two invariant with
respect to SU(2)$\otimes $SU(2), in other words $D^{00}=I_{2,2}$. Global
negativity of partial transpose of state operator, defined as $%
N_{G}^{A_{1}}=2\left\vert I_{2,2}\right\vert $, is a well known entanglement
monotone and quantifies the entanglement of qubit pair $A_{1}A_{2}$.

\subsection{Three-way and two-way correlations}

Determinants of negativity fonts, for partial transposition with respect to
qubit $A_{1}$, of a general three-qubit state%
\begin{equation}
\left\vert \Psi ^{A_{1}A_{2}A_{3}}\right\rangle =\sum_{i_{1}i_{2}}\left(
a_{i_{1}i_{2}0}\left\vert i_{1}i_{2}0\right\rangle
+a_{i_{1}i_{2}1}\left\vert i_{1}i_{2}1\right\rangle \right) ,
\end{equation}%
are defined as $D_{\left( A_{3}\right)
_{i_{3}}}^{00}=a_{00i_{3}}a_{11i_{3}}-a_{10i_{3}}a_{01i_{3}}$ (two-way
fonts), and $D^{00i_{3}}=a_{00i_{3}}a_{11,i_{3}\oplus
1}-a_{10i_{3}}a_{01,i_{3}\oplus 1}$ (three-way fonts), where $i_{3}=0$ or $1$%
. Two-qubit invariants of three-qubit state, corresponding to $I_{2,2}$ are%
\begin{equation}
\left( I_{2}\right) _{A_{3}}^{2,0}=D_{\left( A_{3}\right) _{0}}^{00},\text{
and }\left( I_{2}\right) _{A_{3}}^{0,2}=D_{\left( A_{3}\right) _{1}}^{00},
\end{equation}%
such that $T_{A_{3}}^{+}D_{\left( A_{3}\right) _{0}}^{00}=\left(
D^{000}+D^{001}\right) $. Therefore, the third two-qubit invariant reads as%
\begin{equation}
\left( I_{2}\right) _{A_{3}}^{1,1}=\frac{1}{2}T_{A_{3}}^{+}\left(
I_{2}\right) _{A_{3}}^{2,0}=\frac{\left( D^{000}+D^{001}\right) }{2}.
\end{equation}%
On applying a unitary $U^{A_{3}}$ to $\left\vert \Psi
^{A_{1}A_{2}A_{3}}\right\rangle $, two-qubit invariant $\left( I_{2}\right)
_{A_{3}}^{2,0}$ transforms as%
\begin{equation}
\left( \left( I_{2}\right) _{A_{3}}^{2,0}\right) ^{\prime }=\frac{1}{%
1+\left\vert x\right\vert ^{2}}\left[ \left( I_{2}\right)
_{A_{3}}^{2,0}-2x^{\ast }\left( I_{2}\right) _{A_{3}}^{1,1}+\left( x^{\ast
}\right) ^{2}\left( I_{2}\right) _{A_{3}}^{0,2}\right] .  \label{three}
\end{equation}%
Using Eq. (\ref{inv1}), the three-qubit polynomial invariant is identified as%
\begin{eqnarray}
I_{3,4} &=&\left( I_{2}\right) _{A_{3}}^{2,0}\left( I_{2}\right)
_{A_{3}}^{0,2}-\left( \left( I_{2}\right) _{A_{3}}^{1,1}\right) ^{2}  \notag
\\
&=&D_{\left( A_{3}\right) _{0}}^{00}D_{\left( A_{3}\right) _{1}}^{00}-\left( 
\frac{D^{000}+D^{001}}{2}\right) ^{2}.  \label{i3inv1}
\end{eqnarray}%
This is the invariant which defines three tangle \cite{coff00} through $\tau
_{3,4}=16\left\vert I_{3,4}\right\vert $, an entanglement monotone which
quantifies $3-$way correlations in a three-qubit state. The invariant
corresponding to Eq. (\ref{inv2}) is

\begin{eqnarray}
\mathcal{N}_{3,4}^{A_{1}A_{2}\left( A_{3}\right) } &=&\binom{2}{1}\left\vert
\left( I_{2}\right) _{A_{3}}^{1,1}\right\vert ^{2}+\left\vert \left(
I_{2}\right) _{A_{3}}^{2,0}\right\vert ^{2}+\left\vert \left( I_{2}\right)
_{A_{3}}^{0,2}\right\vert ^{2}  \notag \\
&=&2\left\vert \frac{D^{000}+D^{001}}{2}\right\vert ^{2}+\left\vert
D_{\left( A_{3}\right) _{0}}^{00}\right\vert ^{2}+\left\vert D_{\left(
A_{3}\right) _{1}}^{00}\right\vert ^{2}.  \label{i3inv2}
\end{eqnarray}%
The difference%
\begin{equation}
4\left( \mathcal{N}_{3,4}^{A_{1}A_{2}\left( A_{3}\right) }-2\left\vert
I_{3,4}\right\vert \right) =\left( \tau _{2,2}^{A_{1}A_{2}}\right) ^{2},
\label{i3inv4a1a2}
\end{equation}%
defines $\tau _{2,2}^{A_{1}A_{2}}$, which determines the amount of two-way
correlations in state $\rho ^{A_{1}A_{2}}=$tr$_{A_{3}}\left( \left\vert \Psi
^{A_{1}A_{2}A_{3}}\right\rangle \left\langle \Psi
^{A_{1}A_{2}A_{3}}\right\vert \right) $.

Using the same logic one finds that%
\begin{equation}
\mathcal{N}_{3,4}^{A_{1}A_{3}\left( A_{2}\right) }=\left( 2\left\vert \frac{%
D^{000}-D^{001}}{2}\right\vert ^{2}+\left\vert D_{\left( A_{2}\right)
_{0}}^{00}\right\vert ^{2}+\left\vert D_{\left( A_{2}\right)
_{1}}^{00}\right\vert ^{2}\right) ,  \label{i3inv2a1a3}
\end{equation}%
and%
\begin{equation}
4\left( \mathcal{N}_{3,4}^{A_{1}A_{3}\left( A_{2}\right) }-2\left\vert
I_{3,4}\right\vert \right) =\left( \tau _{2,2}^{A_{1}A_{3}}\right) ^{2},
\label{i3inv4a1a3}
\end{equation}%
where $\tau _{2}^{A_{1}A_{3}}$ quantifies the amount of two-way correlations
in state $\rho ^{A_{1}A_{3}}=$tr$_{A_{2}}\left( \left\vert \Psi
^{A_{1}A_{2}A_{3}}\right\rangle \left\langle \Psi
^{A_{1}A_{2}A_{3}}\right\vert \right) $. The entanglement of qubit $A_{1}$
with qubits $A_{2}$ and $A_{3}$ due to two-way and three-way correlations is
quantified by 
\begin{eqnarray}
\mathcal{N}_{3,4}^{A_{1}} &=&4\left( \mathcal{N}_{3,4}^{A_{1}A_{2}\left(
A_{3}\right) }+\mathcal{N}_{3,4}^{A_{1}A_{3}\left( A_{2}\right) }\right)  
\notag \\
&=&4\left[ \left\vert D^{000}\right\vert ^{2}+\left\vert D^{001}\right\vert
^{2}+\left\vert D_{\left( A_{3}\right) _{0}}^{00}\right\vert ^{2}+\left\vert
D_{\left( A_{3}\right) _{1}}^{00}\right\vert ^{2}\right.   \notag \\
&&\left. +\left\vert D_{\left( A_{2}\right) _{0}}^{00}\right\vert
^{2}+\left\vert D_{\left( A_{2}\right) _{1}}^{00}\right\vert ^{2}\right] .
\label{i3inv}
\end{eqnarray}%
The invariant $\mathcal{N}_{3,4}^{A_{1}}$ is equal to the square of global
negativity of three qubit pure state with respect to qubit $A_{1}$. The
difference 
\begin{equation}
\mathcal{N}_{3,4}^{A_{1}}-\tau _{3,4}=\left( \tau _{2,2}^{A_{1}A_{2}}\right)
^{2}+\left( \tau _{2,2}^{A_{1}A_{3}}\right) ^{2}  \label{i3equality}
\end{equation}%
determines the sum of pairwise entanglement of qubit pairs $A_{1}A_{2}$ and $%
A_{1}A_{3}$. Equalities of Eqs. (\ref{i3inv4a1a2}, \ref{i3inv4a1a3} and \ref%
{i3equality}) are relations between two-way and three-way non-local quantum
correlations in $\left\vert \Psi ^{A_{1}A_{2}A_{3}}\right\rangle $ with
relevant invariants known in terms of determinants of negativity fonts. Eq. (%
\ref{i3equality}) is the analog of CKW inequality \cite{coff00} for three
qubits which states that%
\begin{equation}
\mathcal{N}_{3,4}^{A_{1}}\geq \left( C^{A_{1}A_{2}}\right) ^{2}+\left(
C^{A_{1}A_{3}}\right) ^{2},
\end{equation}%
where squared concurrence, $\left( C^{A_{i}A_{j}}\right) ^{2}$, \cite{hill97}
is a calculable measure of bipartite entanglement of qubits $A_{i}$ and $%
A_{j}$ in the reduced two-qubit state. The concurrence of a two-qubit state $%
\rho $ is defined as%
\begin{equation}
C\left( \rho \right) =\max \left( 0,\sqrt{\lambda _{1}}-\sqrt{\lambda _{2}}-%
\sqrt{\lambda _{3}}-\sqrt{\lambda _{4}}\right) ,
\end{equation}%
where $\lambda _{1}\geq \lambda _{2}\geq \lambda _{3}\geq \lambda _{4}$ are
the eigenvalues of non Hermitian matrix $\widetilde{\rho }\rho $ with $%
\widetilde{\rho }$ $=\left( \sigma _{y}\otimes \sigma _{y}\right) \rho
^{\ast }\left( \sigma _{y}\otimes \sigma _{y}\right) $. Here $\ast $\
denotes complex conjugation in the standard basis and $\sigma _{y}$ is the
Pauli matrix. We have verified that for reduced states obtained from a three
qubit general state 
\begin{equation}
\tau _{2,2}^{A_{i}A_{j}}\geq C^{A_{i}A_{j}}\left( \rho ^{A_{i}A_{j}}\right) .
\end{equation}%
For three qubits, Sudbury \cite{sudb01} lists four algebraically independent
invariants of degree $\leq 4,$ that is $\left\vert \left\langle \Psi
\right\vert \left. \Psi \right\rangle \right\vert ^{2}$, $tr\left( \rho
_{A_{1}}^{2}\right) $, $tr\left( \rho _{A_{2}}^{2}\right) $, $tr\left( \rho
_{A_{3}}^{2}\right) .$ Here $\rho _{A_{i}}$ is the state operator of qubit $%
A_{i}$ which is obtained by tracing over two remaining qubits of the
three-qubit state, for instance $\rho _{A_{1}}=tr_{A_{2}A_{3}}\left(
\left\vert \Psi ^{A_{1}A_{2}A_{3}}\right\rangle \left\langle \Psi
^{A_{1}A_{2}A_{3}}\right\vert \right) $. Invariant of Eq. (\ref{i3inv2}) is
a degree four invariant which can be rewritten as%
\begin{equation}
\mathcal{N}_{3,4}^{A_{1}A_{2}\left( A_{3}\right) }=\frac{1}{4}\left(
1-tr\left( \rho _{A_{1}}^{2}\right) -tr\left( \rho _{A_{2}}^{2}\right)
+tr\left( \rho _{A_{3}}^{2}\right) \right) .
\end{equation}%
Likewise, we have%
\begin{equation}
\mathcal{N}_{3,4}^{A_{1}A_{3}\left( A_{2}\right) }=\frac{1}{4}\left(
1-tr\left( \rho _{A_{1}}^{2}\right) +tr\left( \rho _{A_{2}}^{2}\right)
-tr\left( \rho _{A_{3}}^{2}\right) \right) ,
\end{equation}%
and%
\begin{equation}
\mathcal{N}_{3,4}^{A_{2}A_{3}\left( A_{1}\right) }=\frac{1}{4}\left(
1+tr\left( \rho _{A_{1}}^{2}\right) -tr\left( \rho _{A_{2}}^{2}\right)
-tr\left( \rho _{A_{3}}^{2}\right) \right) .
\end{equation}

\subsection{ Four-way and three-way correlations}

In a general four qubit pure state, written as%
\begin{equation}
\left\vert \Psi ^{A_{1}A_{2}A_{3}A_{4}}\right\rangle
=\sum_{i_{1},i_{2},i_{3}}\left( a_{i_{1}i_{2}i_{3}0}\left\vert
i_{1}i_{2}i_{3}0\right\rangle +a_{i_{1}i_{2}i_{3}1}\left\vert
i_{1}i_{2}i_{3}1\right\rangle \right) ,\quad \left( i_{m}=0,1\right) ,
\end{equation}%
with qubit $A_{1}$ as focus qubit, we identify $D_{\left( A_{3}\right)
_{i_{3}}\left( A_{4}\right)
_{i_{4}}}^{00}=a_{00i_{3}i_{4}}a_{11i_{3}i_{4}}-a_{10i_{3}i_{4}}a_{01i_{3}i_{4}}
$ (two-way), $D_{\left( A_{4}\right)
_{i_{4}}}^{00i_{3}}=a_{00i_{3}i_{4}}a_{11,i_{3}\oplus
1,i_{4}}-a_{10i_{3}i_{4}}a_{01,i_{3}\oplus 1,i_{4}}$ (three-way), $D_{\left(
A_{3}\right) _{i_{3}}}^{00i_{4}}=a_{00i_{3}i_{4}}a_{11i_{3},i_{4}\oplus
1}-a_{10i_{3}i_{4}}a_{01i_{3},i_{4}\oplus 1}$ (three-way), and $%
D^{00i_{3}i_{4}}=a_{00i_{3}i_{4}}a_{11,i_{3}\oplus 1,i_{4}\oplus
1}-a_{10i_{3}i_{4}}a_{01,i_{3}\oplus 1,i_{4}\oplus 1}$- (four-way) as the
determinants of negativity fonts. Using the form of $I_{3,4}$ (Eq. (\ref%
{i3inv1})), one of the three-qubit invariants of four qubit state can be
written as%
\begin{equation}
\left( I_{3}\right) _{A_{4}}^{4,0}=4D_{\left( A_{3}\right) _{0}\left(
A_{4}\right) _{0}}^{00}D_{\left( A_{3}\right) _{1}\left( A_{4}\right)
_{0}}^{00}-\left( D_{\left( A_{4}\right) _{0}}^{000}+D_{\left( A_{4}\right)
_{0}}^{001}\right) ^{2}.  \label{three_4qubit}
\end{equation}%
We have multiplied the invariant by four to facilitate connection with our
earlier work on four qubits \cite{shar13}. Transformation of $\left(
I_{3}\right) _{A_{4}}^{4,0}$ under $U^{A_{4}}$ yields (using Eq. (\ref{kmin}%
))%
\begin{equation}
\left( \left( I_{3}\right) _{A_{4}}^{4,0}\right) ^{\prime }=\frac{1}{\left(
1+\left\vert x\right\vert ^{2}\right) ^{2}}\times \sum\limits_{\mu =0}^{4}%
\binom{4}{\mu }\left( -x^{\ast }\right) ^{\mu }\left( I_{3}\right)
_{A_{4}}^{4-\mu ,\mu }.  \label{4min}
\end{equation}%
Exact expressions (Eqs. (\ref{four31}, \ref{four22}, \ref{four13}, \ref%
{four04})) for additional degree four three-qubit invariants of four qubit
state coefficients, obtained by successive applications of index raising
operator (defined as in Eqs. \ref{rone}, and \ref{rtwo}) to $\left(
I_{3}\right) _{A_{4}}^{4,0}$ are given in Appendix A.

The unitary that makes $\left( \left( I_{3}\right) _{A_{4}}^{4,0}\right)
^{\prime }=0$, is determined by the condition 
\begin{equation}
P(x^{\ast })=\left( I_{3}\right) _{A_{4}}^{4,0}-4x^{\ast }\left(
I_{3}\right) _{A_{4}}^{3,1}+6\left( x^{\ast }\right) ^{2}\left( I_{3}\right)
_{A_{4}}^{2,2}-4\left( x^{\ast }\right) ^{3}\left( I_{3}\right)
_{A_{4}}^{1,3}+\left( x^{\ast }\right) ^{4}\left( I_{3}\right)
_{A_{4}}^{0,4}=0  \label{4condition}
\end{equation}%
which is a degree $4$ polynomial in variable $x^{\ast }$ with three-qubit
invariants as coefficients. Degree eight invariant that detects genuine
four-way entanglement of a four qubit state, constructed by using Eq. (\ref%
{inv1}) for four qubits, is%
\begin{equation}
I_{4,8}=3\left( \left( I_{3}\right) _{A_{4}}^{2,2}\right) ^{2}+\left(
I_{3}\right) _{A_{4}}^{4,0}\left( I_{3}\right) _{A_{4}}^{0,4}-4\left(
I_{3}\right) _{A_{4}}^{3,1}\left( I_{3}\right) _{A_{4}}^{1,3}.
\label{i4inv1}
\end{equation}%
A comparison of Eq. (\ref{4condition}) with Eq. (22) of ref. \cite{luqu03}
shows that our three-qubit invariants, $\left( I_{3}\right)
_{A_{4}}^{4-m,m}\left( m=0\text{ to }4\right) ,$ correspond to coefficients c%
$_{i}\left( i=0\text{ to }4\right) $ of ref. \cite{luqu03}. As such the
correspondence between invariants obtained from (\ref{4condition}) and
invariants presented in refs. \cite{luqu03} and \cite{holw14} is easily
established. Our invariant $I_{4,8}$ corresponds to invariant S of ref. \cite%
{luqu03}. Exact expressions for degree 12 and degree 24 invariants obtained
from Eq. (\ref{4condition}) have been given in ref. \cite{shar13}. Invariant 
$I_{4,8}$ is a function of terms that involve either the products of
invariants that detect $3-$way correlations or involve determinants of
four-way negativity fonts. It is non-zero on states having $4-$way
correlations. Four tangle based on $I_{4,8}$ is defined as%
\begin{equation}
\tau _{4,4}=4\left\vert \left( 12I_{4,8}\right) ^{\frac{1}{2}}\right\vert ,
\label{i4tangle}
\end{equation}%
and calculated in ref. \cite{shar13} for maximally entangled states and
representatives of four qubit states of ref. \cite{vers02}.

By using Eq. (\ref{inv2}) for four qubits, we obtain the invariant%
\begin{equation}
\left( \mathcal{N}_{4,8}\right) ^{A_{1}A_{2}A_{3}\left( A_{4}\right)
}=\sum_{m=0}^{4}\binom{4}{m}\left\vert \left( I_{3}\right)
_{A_{4}}^{4-m,m}\right\vert ^{2}\text{.}  \label{i4inv2}
\end{equation}%
The difference $\left( \mathcal{N}_{4,8}\right) ^{A_{1}A_{2}A_{3}\left(
A_{4}\right) }-\left\vert 2\left( I_{4,8}\right) \right\vert $, determines
the three-way correlations in the reduced state, $\rho ^{A_{1}A_{2}A_{3}}=$
tr$_{A_{4}}\left( \left\vert \Psi ^{A_{1}A_{2}A_{3}A_{4}}\right\rangle
\left\langle \Psi ^{A_{1}A_{2}A_{3}A_{4}}\right\vert \right) $. Likewise,
one can construct the invariants $\left( \mathcal{N}_{4,8}\right)
^{A_{1}A_{2}A_{4}\left( A_{3}\right) }$ and $\left( \mathcal{N}_{4,8}\right)
^{A_{1}A_{3}A_{4}\left( A_{2}\right) }$. We define a measure of three-way
correlations for $A_{1}A_{s}A_{t}$ in four qubit state to be%
\begin{equation}
\left( \tau _{3,4}^{A_{1}A_{s}A_{t}}\right) ^{2}=32\left( \left( \mathcal{N}%
_{4,8}\right) ^{A_{1}A_{s}A_{t}(A_{l})}-\left\vert 2\left( I_{4,8}\right)
\right\vert \right) ,  \label{i4inv4}
\end{equation}%
Four-qubit invariant that quantifies the sum of $3-$way and $4-$way
correlations reads as%
\begin{equation}
\mathcal{N}_{4,8}^{A_{1}}=32\left( \left( \mathcal{N}_{4,8}\right)
^{A_{1}A_{2}A_{3}\left( A_{4}\right) }+\left( \mathcal{N}_{4,8}\right)
^{A_{1}A_{2}A_{4}\left( A_{3}\right) }+\left( \mathcal{N}_{4,8}\right)
^{A_{1}A_{3}A_{4}\left( A_{2}\right) }\right) .  \label{i4inv3}
\end{equation}%
Combining these results and using proper normalization, the difference $%
\mathcal{N}_{4,8}^{A_{1}}-\left( \tau _{4,4}\right) ^{2}$ determines the
amount of total three-way correlations in the four qubit state and satisfies
the equality%
\begin{equation}
\mathcal{N}_{4,8}^{A_{1}}-\left( \tau _{4,4}\right) ^{2}=\left( \left( \tau
_{3,4}^{A_{1}A_{2}A_{3})}\right) ^{2}+\left( \tau
_{3,4}^{A_{1}A_{2}A_{4}}\right) ^{2}+\left( \tau
_{3,4}^{A_{1}A_{3}A_{4}}\right) ^{2}\right) \text{.}  \label{i4equality}
\end{equation}

\subsection{Five-way and Four-way correlations}

The form of four-qubit invariant $\left( I_{4}\right) _{A_{5}}^{8,0}$, which
is a function of state coefficients of five qubit state%
\begin{equation}
\left\vert \Psi ^{A_{1}A_{2}A_{3}A_{4}A_{5}}\right\rangle
=\sum_{i_{1}i_{2}i_{3}i_{4}}\left( a_{i_{1}i_{2}i_{3}i_{4}0}\left\vert
i_{1}i_{2}i_{3}i_{4}0\right\rangle +a_{i_{1}i_{2}i_{3}i_{4}1}\left\vert
i_{1}i_{2}i_{3}i_{4}1\right\rangle \right) ,
\end{equation}%
is known in terms of three-qubit invariants from Eq. (\ref{i4inv1}) and
reads as%
\begin{eqnarray}
\left( I_{4}\right) _{A_{5}}^{8,0} &=&3\left( \left( I_{3}\right)
_{A_{4}}^{2,2}\left( I_{3}\right) _{A_{4}}^{2,2}\right)
_{A_{5}}^{8,0}+\left( \left( I_{3}\right) _{A_{4}}^{4,0}\left( I_{3}\right)
_{A_{4}}^{0,4}\right) _{A_{5}}^{8,0}  \notag \\
&&-4\left( \left( I_{3}\right) _{A_{4}}^{3,1}\left( I_{3}\right)
_{A_{4}}^{1,3}\right) _{A_{5}}^{8,0}\text{.}
\end{eqnarray}%
It transforms under $U^{A_{5}}$ as%
\begin{equation}
\left( \left( I_{4}\right) _{A_{5}}^{8-\mu ,\mu }\right) ^{\prime }=\frac{1}{%
\left( 1+\left\vert x\right\vert ^{2}\right) ^{4}}\sum\limits_{\mu =0}^{8}%
\binom{8}{\mu }\left( -x^{\ast }\right) ^{\mu }\left( I_{4}\right)
_{A_{5}}^{8-\mu ,\mu },
\end{equation}%
and yields the invariant (corresponding to Eq. (\ref{inv1})),%
\begin{eqnarray}
I_{5,16} &=&\left( I_{4}\right) _{A_{5}}^{8,0}\left( I_{4}\right)
_{A_{5}}^{0,8}-8\left( I_{4}\right) _{A_{5}}^{7,1}\left( I_{4}\right)
_{A_{5}}^{1,7}+28\left( I_{4}\right) _{A_{5}}^{6,2}\left( I_{4}\right)
_{A_{5}}^{2,6}  \notag \\
&&-56\left( I_{4}\right) _{A_{5}}^{5,3}\left( I_{4}\right)
_{A_{5}}^{3,5}+35\left( \left( I_{4}\right) _{A_{5}}^{4,4}\right) ^{2},
\label{i5inv1}
\end{eqnarray}%
and (Eq. (\ref{inv2}) for five qubits),%
\begin{equation}
\left( \mathcal{N}_{5,16}\right) ^{A_{1}A_{2}A_{3}A_{4}\left( A_{5}\right)
}=\sum_{m=0}^{8}\binom{8}{m}\left\vert \left( I_{4}\right) _{A_{5}}^{8-\mu
,\mu }\right\vert ^{2}.  \label{i5inv2}
\end{equation}%
In general, for a particular selection of four qubits, we can obtain%
\begin{equation}
\left( \mathcal{N}_{5,16}\right) ^{A_{1}A_{s}A_{t}A_{p}\left( A_{q}\right)
}=\sum_{m=0}^{8}\binom{8}{m}\left\vert \left( I_{4}\right) _{A_{q}}^{8-\mu
,\mu }\right\vert ^{2},
\end{equation}%
where four-qubit invariants $\left( I_{4}\right) _{A_{q}}^{8-m,m}$ ($0\leq
m\leq 8$) are constructed for the selection ($A_{1}A_{s}A_{t}A_{p}$) at
hand.\ As before, being a function of all possible four-qubit invariants of
a five qubit state, the invariant $\left( \mathcal{N}_{5,16}\right) ^{A_{1}}$
defined as the sum%
\begin{eqnarray}
\left( \mathcal{N}_{5,16}\right) ^{A_{1}} &=&C_{5}\left[ \left( \mathcal{N}%
_{5,16}\right) ^{A_{1}A_{2}A_{3}A_{4}\left( A_{5}\right) }+\left( \mathcal{N}%
_{5,16}\right) ^{A_{1}A_{2}A_{3}A_{5}\left( A_{4}\right) }\right.   \notag \\
&&\left. +\left( \mathcal{N}_{5,16}\right) ^{A_{1}A_{2}A_{4}A_{5}\left(
A_{3}\right) }+\left( \mathcal{N}_{5,16}\right) ^{A_{1}A_{3}A_{4}A_{5}\left(
A_{2}\right) }\right] ,  \label{i5inv3}
\end{eqnarray}%
quantifies four-way and five-way correlations. Constant $C_{5}$ is chosen
such that $\left( \mathcal{N}_{5,16}\right) ^{A_{1}}=1,$ on a five qubit GHZ
state. Invariant of Eq. (\ref{i5inv1}) $I_{5,16}$ is non-zero on states
having $5-$way entanglement. The difference, 
\begin{equation}
\left( \tau _{4,4}^{A_{1}A_{s}A_{t}A_{p}}\right) ^{4}=C_{5}\left( \left( 
\mathcal{N}_{5,16}\right) ^{A_{1}A_{s}A_{t}A_{p}\left( A_{q}\right)
}-\left\vert 2I_{5,16}\right\vert \right) ,  \label{i5inv4}
\end{equation}%
determines the four-way correlations in the reduced state $\rho
^{A_{1}A_{s}A_{t}A_{p}}=$tr$_{A_{q}}\left( \rho
^{A_{1}A_{s}A_{t}A_{p}A_{q}}\right) $. The four-qubit invariant $\tau
_{4,4}^{A_{1}A_{s}A_{t}A_{p}}$ is a function of four-qubit invariants $%
\left( I_{4,8}\right) _{A_{q}}^{8-m,m}$ $\left( m=0\text{ to }8\right) $ of
five qubit state. We define a five tangle based on $I_{5,16}$ as%
\begin{equation}
\tau _{5,4}=\left\vert C_{5}8I_{5,16}\right\vert ^{\frac{1}{4}}.
\end{equation}%
The invariants constructed as above, satisfy the equality%
\begin{equation}
\left( \mathcal{N}_{5,16}\right) ^{A_{1}}-\left( \tau _{5,4}\right)
^{4}=\left( 
\begin{array}{c}
\left( \tau _{4,4}^{A_{1}A_{2}A_{3}A_{4}}\right) ^{4}+\left( \tau
_{4,4}^{A_{1}A_{2}A_{3}A_{5}}\right) ^{4} \\ 
+\left( \tau _{4,4}^{A_{1}A_{2}A_{4}A_{5}}\right) ^{4}+\left( \tau
_{4,4}^{A_{1}A_{3}A_{4}A_{5}}\right) ^{4}%
\end{array}%
\right) .  \label{i5equality}
\end{equation}

The process can be continued on to obtain invariants for desired value of $N$%
. While for few qubit systems the process yields exact analytical
expressions for polynomial invariants in terms of functions of state
coefficients, for larger systems it can be implemented, numerically.

We point out that when negativity of partial transpose of a two qubit pure
state is the starting point, then modulus of $N$ qubit unitary invariant $%
I_{N,k}$ of degree $k=2^{N-1}$ ($N\geq 2$) quantifies genuine $N-$way
entanglement in a system of $N$ qubits. Invariant $I_{N,k}$ is non-zero on
states having genuine $N-$body correlations and is zero on all states not
having genuine $N-$body correlations independent of whether the state has $%
K- $body correlations or not, where $K<N$. For a pure state, it can be
understood as the residual entanglement not accounted for by the sum of
two-way, three-way,...,$\left( N-1\right) $ way entanglement between
individual qubits. The invariant $\left( \mathcal{N}_{N,k}\right) ^{A_{1}}$
measures the sum of $N-$way and $\left( N-1\right) -$way correlations of
qubit $A_{1}$ with the rest of the system. For a given choice of $\left(
N-1\right) $ qubits, an invariant that is the sum of $\left( N-1\right) -$%
way quantum correlations amongst the focus set and $N-$way correlations is
also found. These results reduce the question of monogamy of quantum
correlations in a multiqubit system to establishing the relation between an
independent measure of $\left( N-1\right) $ qubit correlations in reduced $%
\left( N-1\right) $ qubit state and the corresponding pure state invariants.

\section{Conclusions}

Local unitary invariance is the basic principle used to formulate an
inductive process that generates a chain of polynomial invariants of state
coefficients starting from a known subsystem invariant of a multipartite
system. Our approach of selective sequential construction allows for
construction of physically meaningful invariant functions for a large system
in terms of known properties of subsystems. The process is applied to obtain
the chain of invariants based on global negativity which is an entanglement
monotone for a two qubit pure state. The resulting degree $k$ invariant $%
I_{N,k}$ ($k=2^{N-1}$) detects entanglement due to $N-$way non-local
correlations in an $N$ qubit state and is used to define $N-$tangle, a
quantifier of $N-$way correlations. Other interesting polynomial invariants
that quantify the sum of $N-$way and $\left( N-1\right) -$way correlations
in a pure state and $\left( N-1\right) -$way correlations in reduced states
have, also, been obtained. Invariants for mixed states can be constructed
through convex roof extension.

Since the form of $N-$qubit invariants is directly linked to the underlying
structure of the composite system state, it can throw light on the
suitability of a given state for a specific information processing task.
Polynomial invariants generated by the process can help solve the related
problems of local unitary equivalence and classification of states. The
process is quite general and easily extendable to $d-$dimensional
subsystems. We verify that specific invariants of $K<N$ qubits are a
restriction on invariants on $N$ qubits.

Recently, Li and Li \cite{li12} have used invariance of the rank of the
coefficient matrix associated with the pure state of $N$ qubits to classify $%
N$ qubit states. B. Liu et al. \cite{liu12} exploit the local symmetries of
the states, to classify general multipartite pure states. Polynomial
invariants that identify the nature of correlations in a state are useful to
apply classification criteria based on type of correlations present in the
state \cite{shar12}.

Mathematically, polynomial invariants of $N$ qubit system expressed in terms
of determinants of negativity fonts are relations between invariants of a
matrix in Hilbert space of dimension $2^{N}$ and negative eigenvalues of
submatrices in $4\times 4$ subspaces. We may point out that the inductive
process yields exact relations between eigenvalues of a large matrix and its
submatrices which is an interesting mathematical result.

This work is supported by Faep Uel, Funda\c{c}\~{a}o Araucaria PR and CNPq,
Brazil.

\appendix

\section{Set of degree four three-qubit invariants of a four qubit state}

Elements in the set of degree four three-qubit invariants of four qubit
state coefficients are related to each other through index raising operator
defined in Eqs. \ref{rone} and \ref{rtwo}. Starting with three-qubit
invariant (Eq. \ref{three_4qubit})%
\begin{equation}
\left( I_{3}\right) _{A_{4}}^{4,0}=4D_{\left( A_{3}\right) _{1}\left(
A_{4}\right) _{0}}^{00}D_{\left( A_{3}\right) _{0}\left( A_{4}\right)
_{0}}^{00}-\left( D_{\left( A_{4}\right) _{0}}^{000}+D_{\left( A_{4}\right)
_{0}}^{001}\right) ^{2},
\end{equation}%
we obtain%
\begin{eqnarray}
\left( I_{3}\right) _{A_{4}}^{3,1} &=&\frac{1}{4}T_{A_{4}}^{+}\left(
I_{3}\right) _{A_{4}}^{4,0}  \notag \\
&=&D_{\left( A_{3}\right) _{1}\left( A_{4}\right) _{0}}^{00}\left( D_{\left(
A_{3}\right) _{0}}^{000}+D_{\left( A_{3}\right) _{0}}^{001}\right)
+D_{\left( A_{3}\right) _{0}\left( A_{4}\right) _{0}}^{00}\left( D_{\left(
A_{3}\right) _{1}}^{000}+D_{\left( A_{3}\right) _{1}}^{001}\right)   \notag
\\
&&-\frac{1}{2}\left( D_{\left( A_{4}\right) _{0}}^{000}+D_{\left(
A_{4}\right) _{0}}^{001}\right) \left(
D^{0000}+D^{0001}+D^{0010}+D^{0011}\right) ,  \label{four31}
\end{eqnarray}%
\begin{eqnarray}
\left( I_{3}\right) _{A_{4}}^{2,2} &=&\frac{1}{3}\left( T_{A_{4}}^{+}\right)
^{2}\left( I_{3}\right) _{A_{4}}^{4,0}  \notag \\
&=&\frac{2}{3}\left( D_{\left( A_{3}\right) _{1}}^{000}+D_{\left(
A_{3}\right) _{1}}^{001}\right) \left( D_{\left( A_{3}\right)
_{0}}^{000}+D_{\left( A_{3}\right) _{0}}^{001}\right)   \notag \\
&&+\frac{2}{3}\left( D_{\left( A_{3}\right) _{1}\left( A_{4}\right)
_{0}}^{00}D_{\left( A_{3}\right) _{0}\left( A_{4}\right)
_{1}}^{00}+D_{\left( A_{3}\right) _{0}\left( A_{4}\right)
_{0}}^{00}D_{\left( A_{3}\right) _{1}\left( A_{4}\right) _{1}}^{00}\right)  
\notag \\
&&-\frac{1}{6}\left( D^{0000}+D^{0001}+D^{0010}+D^{0011}\right) ^{2}  \notag
\\
&&-\frac{1}{3}\left( D_{\left( A_{4}\right) _{0}}^{000}+D_{\left(
A_{4}\right) _{0}}^{001}\right) \left( D_{\left( A_{4}\right)
_{1}}^{000}+D_{\left( A_{4}\right) _{1}}^{001}\right)   \label{four22}
\end{eqnarray}%
\begin{eqnarray}
\left( I_{3}\right) _{A_{4}}^{1,3} &=&\frac{1}{24}\left(
T_{A_{4}}^{+}\right) ^{3}\left( I_{3}\right) _{A_{4}}^{4,0}  \notag \\
&=&D_{\left( A_{3}\right) _{1}\left( A_{4}\right) _{1}}^{00}\left( D_{\left(
A_{3}\right) _{0}}^{000}+D_{\left( A_{3}\right) _{0}}^{001}\right) +\left(
D_{\left( A_{3}\right) _{1}}^{000}+D_{\left( A_{3}\right) _{1}}^{001}\right)
D_{\left( A_{3}\right) _{0}\left( A_{4}\right) _{1}}^{00}  \notag \\
&&-\frac{1}{2}\left( D^{0000}+D^{0001}+D^{0010}+D^{0011}\right) \left(
D_{\left( A_{4}\right) _{1}}^{000}+D_{\left( A_{4}\right) _{1}}^{001}\right)
,  \label{four13}
\end{eqnarray}%
and%
\begin{eqnarray}
\left( I_{3}\right) _{A_{4}}^{0,4} &=&\frac{1}{24}\left(
T_{A_{4}}^{+}\right) ^{4}\left( I_{3}\right) _{A_{4}}^{4,0}  \notag \\
&=&4D_{\left( A_{3}\right) _{1}\left( A_{4}\right) _{1}}^{00}D_{\left(
A_{3}\right) _{0}\left( A_{4}\right) _{1}}^{00}-\left( D_{\left(
A_{4}\right) _{1}}^{000}+D_{\left( A_{4}\right) _{1}}^{001}\right) ^{2}.
\label{four04}
\end{eqnarray}

\section{Binary forms and Invariants of Polynomials by standard method of
transvection}

The process outlined in section (II) results in binary forms necessary to
obtain $N$ qubit invariants. The coefficients of the forms are $\left(
N-1\right) $ qubit invariants. More over, these invariants are known in
terms of determinants of $K-$way negativity fonts ($2\leq K\leq N$), having
physical meaning as possible sources of $K-$body correlations. In this
appendix, we give the formulae relevant to obtaining the two invariants
given in Eqs. ((\ref{inv1}) and (\ref{inv2})) from Eq. (\ref{kmin}). Algebra
necessary for calculating the complete set of invariants for $N-$qubits in
terms of determinants of negativity fonts, is found in refs. \cite%
{davi93,olve99,jans11}.

A homogeneous binary form, $f\left( x,y\right) ,$ of degree $k$ in variables 
$x$ and $y$ is defined as 
\begin{equation}
f\left( x,y\right) =\sum\limits_{m=0}^{k}a_{k-m}x^{k-m}y^{m},
\end{equation}%
where the coefficient $a_{k-m}=\frac{1}{m!\left( k-m\right) !}\left( \frac{%
\partial }{\partial x}\right) ^{k-m}\left( \frac{\partial }{\partial y}%
\right) ^{m}f\left( x,y\right) $ for $0\leq m\leq k$. We can identify the
form $f\left( x,1\right) $ with the polynomial%
\begin{equation}
P_{k}\left( x\right) =\sum\limits_{m=0}^{k}a_{k-m}x^{k-m},
\end{equation}%
which gives a one-to-one correspondence between binary forms of degree $k$
and polynomials of degree (at most) $k$ described by $f\left( x,y\right) $ = 
$f\left( x,1\right) $. On application of a definite set of transformations
upon variables $x,y$ the form $f$ transforms to $f^{\prime }$. Let there be
associated with $f$ some definite quantity $I$ such that when the
corresponding quantity $I^{\prime }$ is constructed for the transformed
function $f^{\prime }$ the equality $I^{\prime }=MI$ holds. If $M$ depends
only upon the transformations, that is, is free from any relationship with $%
f $, then $I$ is called an invariant of $f$ under the transformations of the
set.

A differential operation called transvection is the most fundamental process
of binary invariant theory. All invariants and covariants of a form or a set
of forms can be derived by this process. Let $f\left( x_{1},y_{1}\right)
=\sum\limits_{m=0}^{k}a_{k-m}x_{1}^{k-m}y_{1}^{m}$ be a binary form of
degree $k$ and $g\left( x_{2},y_{2}\right)
=\sum\limits_{m=0}^{n}b_{n-m}x_{2}^{n-m}y_{2}^{m},$ a binary form of degree $%
n$. The $r^{th}$ transvectant of $f\left( x_{1},y_{1}\right) $ and $g\left(
x_{2},y_{2}\right) $ with $r\leq \min \left\{ k,n\right\} $, abbreviated as $%
(f,g)^{r}$ can be calculated by using the formula%
\begin{equation}
(f,g)^{r}=\frac{\left( n-r\right) !\left( k-r\right) !}{n!k!}%
\sum_{s=0}^{r}\left( -1\right) ^{s}\binom{r}{s}\frac{\partial ^{r}f\left(
x_{1},y_{1}\right) }{\partial ^{r-s}x_{1}\partial ^{s}y_{1}}\frac{\partial
^{r}g\left( x_{2},y_{2}\right) }{\partial ^{s}x_{2}\partial ^{r-s}y_{2}}%
\text{.}
\end{equation}%
Consider the specific case of binary form%
\begin{equation}
f\left( x_{1},y_{1}\right) =\sum\limits_{\mu _{1}=0}^{k}\binom{k}{\mu _{1}}%
\left( x_{1}\right) ^{\mu _{1}}\left( y_{1}\right) ^{k-\mu _{1}}\left(
I_{N-1}\right) _{A_{N}}^{k-\mu _{1},\mu _{1}},
\end{equation}%
where%
\begin{equation*}
\binom{k}{\mu _{1}}\left( I_{N-1}\right) _{A_{N}}^{k-\mu _{1},\mu _{1}}=k!%
\frac{\partial ^{k-\mu _{1}}}{\partial x}\frac{\partial ^{\mu _{1}}}{%
\partial y}f\left( x_{1},y_{1}\right) .
\end{equation*}%
We identify $f\left( x_{1}=-x^{\ast },y_{1}=1\right) $ with $\left( \left(
I_{N-1}\right) _{A_{N}}^{k,0}\right) ^{\prime }\left( \left( 1+\left\vert
x\right\vert ^{2}\right) ^{\frac{k}{2}}\right) $ of Eq. (\ref{kmin}). The $%
k^{th}$ transvectant of $f\left( x_{1},y_{1}\right) $ with itself that is $%
(f,f)^{k}$ is an invariant of binary form. It is given by 
\begin{equation}
(f,f)^{k}=\frac{1}{k!k!}\sum_{s=0}^{k}\left( -1\right) ^{s}\binom{k}{s}\frac{%
\partial ^{k}f\left( x_{1},y_{1}\right) }{\partial ^{k-s}x_{1}\partial
^{s}y_{1}}\frac{\partial ^{k}f\left( x_{1},y_{1}\right) }{\partial
^{s}x_{1}\partial ^{k-s}y_{1}}
\end{equation}%
Defining $I_{N,2k}$ =$\frac{1}{2}(f,f)^{k}$, we obtain 
\begin{equation}
I_{N,2k}=\frac{1}{2}\sum_{s=0}^{k}\left( -1\right) ^{s}\binom{k}{s}\left(
I_{N-1}\right) _{A_{N}}^{k-s,s}\left( I_{N-1}\right) _{A_{N}}^{s,k-s},
\end{equation}%
which for $k=4$ reads as%
\begin{equation}
I_{4,8}=\left( I_{N-1}\right) _{A_{N}}^{4,0}\left( I_{N-1}\right)
_{A_{N}}^{0,4}-4\left( I_{N-1}\right) _{A_{N}}^{3,1}\left( I_{N-1}\right)
_{A_{N}}^{1,3}+3\left( \left( I_{N-1}\right) _{A_{N}}^{2,2}\right) ^{2}.
\end{equation}%
Next consider%
\begin{equation}
g\left( x_{2},y_{2}\right) =\sum\limits_{\mu _{2}=0}^{k}\binom{k}{\mu _{2}}%
\left( x_{2}\right) ^{k-\mu _{2}}\left( y_{2}\right) ^{\mu _{2}}\left(
\left( I_{N-1}\right) _{A_{N}}^{\mu _{2},k-\mu _{2}}\right) ^{\ast },
\end{equation}%
such that $g\left( x_{2}=1,y_{2}=x^{\ast }\right) =\left( 1+\left\vert
x\right\vert ^{2}\right) ^{\frac{k}{2}}\left( \left( \left( I_{N-1}\right)
_{A_{N}}^{0,k}\right) ^{\prime }\right) ^{\ast }$ (from Eq. (\ref{kmin})).
The k$^{th}$ simultaneous transvectant of the forms $f$ and $g$ is 
\begin{eqnarray}
(f\left( x_{1},y_{1}\right) ,g\left( x_{2},y_{2}\right) )^{k} &=&\frac{1}{%
k!k!}\sum_{s=0}^{k}\left( -1\right) ^{s}\binom{k}{s}\frac{\partial
^{k}f\left( x_{1},y_{1}\right) }{\partial ^{k-s}x_{1}\partial ^{s}y_{1}}%
\frac{\partial ^{k}g\left( x_{2},y_{2}\right) }{\partial ^{s}x_{2}\partial
^{k-s}y_{2}}  \notag \\
&=&\sum_{s=0}^{k}\binom{k}{s}\left\vert \left( I_{N-1}\right)
_{A_{N}}^{k-s,s}\right\vert ^{2}.
\end{eqnarray}%
In other words the invariant of Eq. (\ref{inv2}), $\left( \mathcal{N}%
_{N,2k}\right) ^{A_{1}A_{2}...A_{N-1}\left( A_{N}\right) }=(f\left(
x_{1},x_{2}\right) ,g\left( x_{1},x_{2}\right) )^{k}.$ The invariants of a
binary form are a graded algebra, and Gordan \cite{gord68} proved that this
algebra is finitely generated if the base field is the complex numbers.

\end{document}